\documentclass[final,3p,times]{elsarticle}

\usepackage{graphics}

\usepackage{amssymb}

\journal{International Journal of Mass Spectrometry}

\begin{document}

\begin{frontmatter}



\title{Nuclear Masses and Neutron Stars}


\author[1,2]{S. Kreim}
\author[3]{M. Hempel}
\author[4]{D. Lunney}
\author[5]{J. Schaffner-Bielich}

\address[1]{CERN, CH-1211 Geneva, Switzerland}
\address[2]{Max-Planck-Institut f\"ur Kernphysik, Saupfercheckweg 1, 69117 Heidelberg, Germany}
\address[3]{Department of Physics, University of Basel, 4056 Basel, Switzerland}
\address[4]{CSNSM-IN2P3-CNRS, Universit\'e Paris-Sud, 91406 Orsay, France}
\address[5]{Institut f\"ur Theoretische Physik, Goethe Universit\"at, 60438 Frankfurt am Main, Germany}

\begin{abstract}
Precision mass spectrometry of neutron-rich nuclei is of great relevance for astrophysics.  Masses of exotic nuclides impose constraints on models for the nuclear interaction and thus affect the description of the equation of state of nuclear matter, which can be extended to describe neutron-star matter.  With knowledge of the masses of nuclides near shell closures, one can also derive the neutron-star crustal composition.

The Penning-trap mass spectrometer ISOLTRAP at CERN-ISOLDE has recently achieved a breakthrough measuring the mass of $^{82}$Zn, which allowed constraining neutron-star crust composition to deeper layers \cite{Wolf13b}. We perform a more detailed study on the sequence of nuclei in the outer crust of neutron stars with input from different nuclear models to illustrate the sensitivity to masses and the robustness of neutron-star models. The dominant role of the $N=50$ and $N=82$ closed neutron shells for the crustal composition is confirmed.
\end{abstract}


\end{frontmatter}


\section{Introduction}
\label{intro}
Neutron stars (NS) are among the most compact objects known. They are the result of the collapse of massive stars and are composed mainly - but not completely - of neutrons. After their rather violent birth in core-collapse supernovae, they cool quite quickly (via the emission of neutrinos) within a minute and reach a state of (small temperature) equilibrium. The equilibrium composition of a neutron star reflects the surprising but logical result of a very wide range of physics \cite{Chamel08}:  from stable nuclides on the surface, to exotic nuclides in the crust, exotic forms of matter (a series of so-called pasta phases \cite{Lattimer07}) to a core that may even contain unbound quarks. 

A widely-used concept for modeling astrophysical environments at high densities is nuclear matter, a homogeneous, infinite medium composed of protons and neutrons, characterized by a certain equation of state (EOS). Some of the properties of this EOS could be inferred from experimental observations. Neutron stars are the astrophysical objects with the closest resemblance to a nuclear-matter environment. However, in neutron stars, the EOS is probed at extremes of isospin and densities beyond the saturation density of symmetric nuclear matter of $\rho_0�\simeq 0.16$~fm$^{-3}\simeq 2 \times10^{14}$~g/cm$^3$. Therefore the equation of state of nuclear matter has to be extended from symmetric nuclear matter to describe neutron matter, and with the assumption of $\beta$-equilibrium then neutron-star matter. Furthermore, electrons have to be added to maintain electric charge neutrality. A realistic formulation of the EOS is required for a robust neutron-star model, because the EOS has a direct impact on the 
possible combinations of mass and radius, which obey hydrostatic equilibrium in general relativity.

Neutron-star models based on different EOS still differ strongly in their predictions for the properties of a neutron star. Even though there are many neutron stars with masses measured by radio astronomy \cite{Lattimer12} (in this context they are called pulsars, due to their pulsed radio signal), the associated fundamental quantity of a neutron star, the radius, is extremely difficult to extract from astronomical observations, due to their size of only 10 to $20\,$km. However, this field has made significant progress in the last years by theoretical modeling of observations of low-mass X-ray binaries or cooling neutron stars \cite{Steiner12a,Steiner12b,Suleimanov12,Guever13}. A recent breakthrough was the discovery of a relatively heavy neutron star -- the pulsar J1614-2230 with 1.97(4) solar masses -- which ruled out several neutron-star models \cite{Demorest10,Lattimer10}. A very exciting future perspective is gravitational wave astronomy. Recently, a tight correlation between the frequency peak of the 
post-merger gravitational-wave emission and the radius of a 1.6 solar mass neutron star has been demonstrated \cite{Bauswein12}.

Correctly modeling neutron-star matter requires a proper description of nuclear interactions, and especially three-body forces. Alternatively, effective models in the form of density functionals can be used. The theoretical description of the stability of nuclides close to the drip line and the equation of state of neutron-star matter is sensitive to details of the nuclear Hamiltonian. The proper saturation of nuclear matter has been imposed as an almost necessary condition for any nuclear potential \cite{Bethe71}. The exact location of the drip lines along the nuclear chart is of utmost importance to understand the basic concept of nuclear stability and the underlying theory of the nuclear force \cite{Hansen87}. Still today, a consistent description of the nuclear force remains a challenge \cite{Erler12}. 

The so-called rapid and slow neutron-capture processes (r- and s-process) of stellar nucleosynthesis are responsible for the origin of most of the neutron-rich, heavy elements \cite{Schatz13}. To date, the site of a successful r-process nucleosynthesis is still unknown \cite{Arnould07}. A possible site, complementary to the supernova-induced r-process, is the ejection of neutron-star matter by a merger of two neutron star. This allows an robust r-process to occur as the ejected clump vaporizes into the interstellar medium and undergoes nuclear reactions. This scenario could explain at least part of the total enrichment of the heavy r-process elements in the Galaxy \cite{Arnould07,Korobkin12}. The neutron-star crust offers the required ingredients for a successful r-process, because of the presence of neutron-rich material, i.e., with a low electron fraction \cite{Korobkin12}. In the outer crust, the electron fraction is directly given by the charge-to-mass ratio of the heavy 
nuclei. On the other hand, the nuclear binding energy is the decisive quantity for establishing whether or not a certain isotope is present in the neutron-star crust. Precise mass measurements are thus valuable input for the composition of the crust and r-process nucleosynthesis \cite{Chamel08}.

A striking observable related to mass measurements is the fact that crossing a magic proton or neutron number produces a sharp drop in the corresponding one-particle or two-particle separation energy, an effect well established as a signature for magicity. In the light nuclides, the disappearance of this effect has been regarded as evidence for the reduction of the shell gap (the so-called ``shell quenching'' phenomenon) \cite{Sorlin08}, while an enhancement of the effect would be, on the contrary, evidence for the emergence of a new magic number.  The evolution of the magic numbers $N = 50$ and $N = 82$ far from stability affects the equilibrium composition of the neutron-star crust and potentially the predicted elemental abundance. Dramatic changes in nuclear structure far from stability are very challenging for theory and the correct prediction of shell evolution far from stability is perhaps the strongest challenge for the different nuclear models and corresponding nuclear interactions. 

The article highlights some very recent advances in determining the properties of neutron stars through nuclear mass spectrometry. In a first part, the link from nuclear masses to the nuclear-matter EOS will be established, based on state-of-the-art calculations starting from two- and three-body interactions derived within chiral effective field theory (chiral EFT). In a second part, recent constraints on the equilibrium composition of the neutron-star crust will be presented, determined by the mass measurement of $^{82}$Zn \cite{Wolf13b}. The robustness of the determined equilibrium composition to different global mass models used to supplement the experimentally unavailable data will be explored.

\section{Nuclear interaction and the equation of state}
One of the long-standing goals of theoretical nuclear physics is the development of a nuclear Hamiltonian, allowing the description, in a unified microscopic picture, of the properties of all nuclear systems. A direct link to the underlying theory of quantum chromodynamics (QCD) has been difficult to establish due to the highly non-perturbative character of the theory at the nuclear energy scale. The historical approach was to build the so-called ``realistic'' potentials, designed to fit the low-energy nucleon-nucleon scattering data and the properties of few-nucleon systems, such as the binding energy of the deuteron \cite{Bethe71,Jensen95}. The major test of realistic potentials has come from the comparison of their predictions with the properties of light and medium-mass nuclides, through ab initio \cite{Pudliner97} and shell-model \cite{Caurier05} calculations. Apart from the two-body content of nuclear interactions, three-body forces arise at a fundamental level in low-energy nuclear theory due to the 
finite size of the nucleons, the internal structure of which can be virtually excited in a three-body process \cite{Fujita57}. Recently, a new class of realistic potentials has been developed, based on chiral EFT, using as starting point the symmetries of the QCD Lagrangian and only retaining the degrees of freedom relevant for low-energy nuclear theory. Three-body forces arise naturally in the chiral EFT expansion of the Hamiltonian \cite{Epelbaum09}. Most recently, several groups have provided a consistent description of neutron matter based on chiral EFT interactions \cite{Tews13,Sammarruca12,Holt12a,Coraggio13}. 

\begin{figure}
\includegraphics[width=0.7\textwidth]{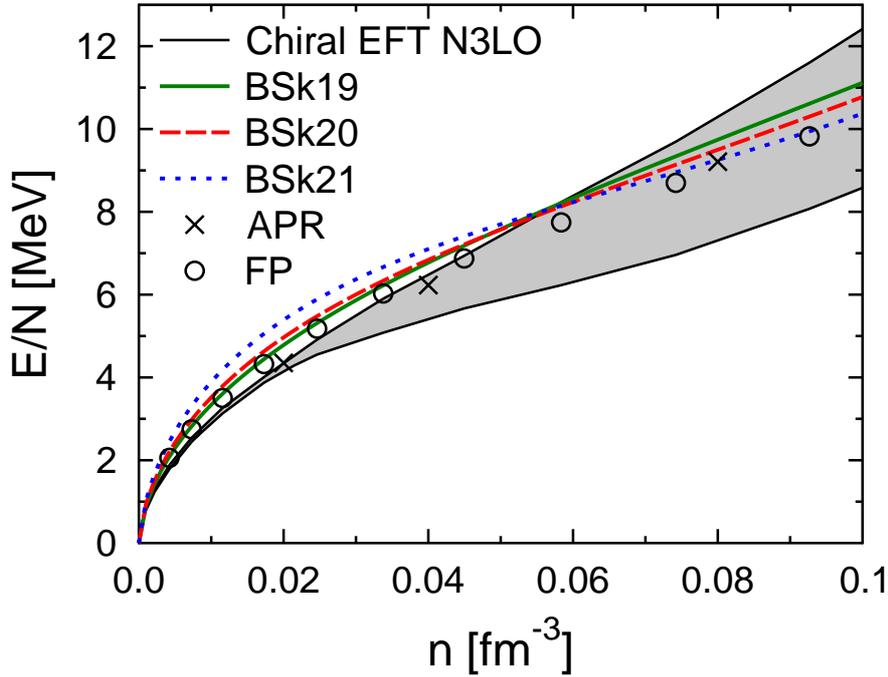}
\caption{\label{fig:nm_eos}Neutron matter energy per nucleon as a function of density for chiral EFT calculations with three-body forces at next-to, next-to, next-to leading order (N3LO) \cite{Tews13}, Skyrme forces BSk19-21 \cite{Pearson12}, the EOS of Friedman and Pandharipande (FP) \cite{Friedman81} and of Akmal, Pandharipande and Ravenhall (APR) \cite{Akmal98}. For details, see text.}
\end{figure}

Only last year, calculations for the calcium isotopic chain with a chiral EFT potential containing three-body forces have successfully described the doubly-magic nucleus $^{48}$Ca starting from the $^{40}$Ca core, which has posed problems for shell-model calculations using exclusively two-body nuclear interactions \cite{Holt12b}. The calcium isotopic chain is an intriguing case of nuclear magicity at neutron number $N$ = 32.  The high energy of the assumed first $2^+$ excited state in $^{52}$Ca suggest the enhancement of the $N$ = 32 subshell gap \cite{Huck85}. Very recent Penning-trap mass measurements of $^{ 51,52}$Ca, performed at TRIUMF with the TITAN experiment, show very large discrepancies from the AME extrapolations \cite{AME03}, suggesting significant structural changes in the region \cite{Gallant12}. Time-of-flight mass measurements at NCSL of scandium isotopes $^{53-55}$Sc hint at a drop in the two-neutron separation energy at $N$ = 32, but the large uncertainties do not allow drawing a definite 
conclusion \cite{Estrade11}. A very recent breakthrough came through the measurement of the masses of $^{53,54}$Ca \cite{Wienholtz12} at ISOLDE \cite{Kugler00} with the ISOLTRAP experiment \cite{Mukherjee08,Kluge13}. The minute yield and short half-life of these isotopes were prohibitive for a Penning trap mass measurement. The masses of $^{53,54}$Ca were thus determined directly with the multi-reflection time-of-flight mass separator (MR-TOF MS) with a precision better than parts-per-million (ppm). (For details on the MR-TOF MS, see \cite{Wolf12,Wolf13a}) Shell-model calculations performed for the calcium isotopic chain with a chiral EFT potential containing three-body forces agree well with the experimental data, as well as with calculations with phenomenological potentials, such as KB3G and GXPF1A \cite{Gallant12}. The chiral EFT potential, including its three-body part, is constrained only by nucleon-nucleon scattering data and the properties of only two light nuclides: $^3$H and $^4$He. These recent 
results support chiral EFT as a promising global approach for nuclear theory.

The equation of state plays a central role in astrophysical observations, one application being NS matter and the structure of NS \cite{Baym71a,Lattimer01,Lattimer04,Lattimer07}. The description of neutron matter with microscopic calculations based on chiral EFT interactions has been tackled recently by Hebeler and co-workers. They were able to constrain the properties of neutron-rich matter below nuclear densities to a much higher degree than is reflected in many commonly used EOS \cite{Hebeler10a,Hebeler10b,Hebeler11}. In their calculations, they include three-body forces to the third order and provide theoretical uncertainties. Just recently, they included also the subleading three-body forces for the first time and all leading four-nucleon (4N) forces. In Fig.~\ref{fig:nm_eos} the neutron matter energy band of their calculation is shown including evaluated theoretical uncertainties. The band includes the use of different nucleon-nucleon potentials, uncertainties in the three-body forces, and in the many-
body calculation, for details see \cite{Tews13}. Their results not only provide constraints for the nuclear EOS as well as neutron-rich matter in astrophysics but also agree with other recent work \cite{Sammarruca12,Holt12a,Coraggio13}. One hope is that these models will be extended to provide also all the necessary mass data for probing the composition of neutron stars. 

Figure \ref{fig:nm_eos} also shows the neutron-matter equations of state (internal energy per nucleon as a function of density) for forces BSk19-21 at subnuclear densities and zero temperature \cite{Pearson12}. Goriely and colleagues have used in their fit for the effective forces BSk19-21 constraints from neutron-matter calculations as well as mass data. They can thus consistently provide not only a description of neutron matter but also of binding energies of all nuclides. However, in their fit they used other realistic neutron matter EOS mainly at supersaturation densities. Two of these EOSs (FP \cite{Friedman81} and APR \cite{Akmal98}) are also shown in Fig.~\ref{fig:nm_eos}. Despite some deviations at lowest densities the different calculations show an good agreement. The neutron matter EOS is generally of great importance for the very neutron-rich neutron star EOS and should help provide a consistent description of the very asymmetric nuclei formed in the crust.

\section{The outer-crust composition of neutron stars in $\beta$-equilibrium}

\begin{figure}
\includegraphics[width=0.5\textwidth]{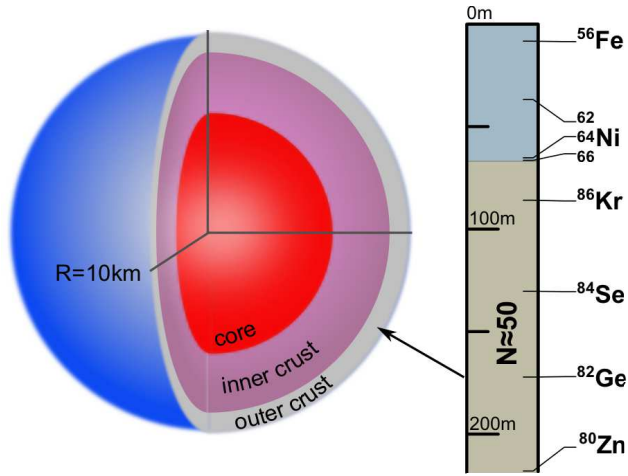}
\caption{\label{fig:structure}The depth profile of a neutron star of 1.4 solar mass and 10\,km radius. The scale on the right indicates the nuclidic composition in $\beta$-equilibrium of the outer crust as determined by measured masses \cite{AME12} and the recent $^{82}$Zn result \cite{Wolf13b}.}
\end{figure}

Neutron stars are the remnants of core-collapse supernova explosions. After having reached $\beta$-equilibrium, their surface contains the nuclides $^{56}$Fe and $^{62}$Ni, which have the highest binding energy per nucleon. Note that $^{56}$Fe has been observed in the accretion disk around neutron stars \cite{Cacket08}. Three distinct regions are thought to compose a neutron star (Fig.~\ref{fig:structure}): a locally homogeneous core and two concentric shells characterized by different inhomogeneous phases forming a solid crust \cite{Baym71a,Baym71b,Negele73,Lattimer04}. The ``outer crust'', consists of a crystal of ionized atoms coexisting with a quantum gas of electrons. At these extreme pressures encountered in the interior of neutron stars, the electrons are squeezed into the nuclides, shifting the equilibrium to heavier and more neutron-rich isotopes. Burrowing into the crust, nuclides become more and more neutron-rich up to the point where neutrons start to drip out. At the so-called neutron-drip 
density (about $4\cdot10^{11}\,$g/cm$^3$), unbound neutrons form a separate additional phase. This marks the transition to the ``inner crust'', a lattice-like structure of nuclear systems immersed in a sea of unbound neutrons and electrons. The similarity of this state to that of a crystal lattice in a sea of conducting electrons gave rise to use the model of the Wigner-Seitz cell (WS approximation) \cite{Wigner33,Wigner34}. The WS cells, with an associated lattice energy, are possibly transformed by further pressure into different exotic shapes, called nuclear pasta \cite{Pais12}, due to a frustration of the system with respect to Coulomb and surface energies. Deeper into the star, the crust dissolves into a uniform liquid of nucleons and leptons and the core is reached.

To determine the crustal composition, one has to assume that the neutron-star crust within the WS approximation is in full thermodynamic equilibrium.  To determine the equilibrium composition of a given cell, the principle of minimization of the Gibbs free energy is used.  For a given pressure, this will depend on the lattice energy, the electron energy and - what is interesting from the point of view of nuclear physics:  the binding energy of the nucleus. The other quantities are from well-known classical (atomic) physics and thermodynamics so the modeling of these systems is relatively robust. To determine the depth profile, the integration of the relativistic hydrostatic Tolman-Oppenheimer-Volkoff (TOV) equations over a pressure column is required \cite{Baym71b}. Such a calculation fixes also the total abundances of all nuclear species in the crust \cite{Hempel07}. Because of nuclear shell effects, the exotic nuclides residing in neutron-star crusts accumulate around the magic neutron number $N$ = 50 and 
$N$ = 82. The sequence of nuclides that occur in the outer crust of nonaccreting cold neutron stars was investigated by Baym, Pethick, and Sutherland (BPS) in what has become a landmark paper in 1971 \cite{Baym71b}. Not all nuclides which are relevant for calculating the equilibrium composition of the outer crust have an experimentally known mass. For these nuclides, the mass is calculated using mass models, which extrapolate semi-empirically from the known masses into unmeasured regions. Baym et al.~used the mass data from the droplet model of Myers and Swiatecki \cite{Myers66}. 

Using known masses and state-of-the-art nuclear mass models, the equilibrium composition of the outer crust had already been robustly determined to a depth of about 212\,m for a canonical neutron star of 1.4 solar mass and 10\,km radius \cite{Ruster06,Chamel11}. Since the different mass models used predict different equilibrium compositions, the crustal composition can only be pinned down by high-precision mass measurements. The most exotic nuclides predicted in the crust at the $N$ = 50 shell, $^{82}$Zn has been determined with the Penning-trap mass spectrometer ISOLTRAP \cite{Wolf13b}. The calculations performed there were restricted to the three most recent Brussels-Montreal mass tables HFB-19, HFB-20, and HFB-21 since they are also constrained to reproduce the EOS of neutron matter from calculations with realistic nucleon-nucleon potentials. Figure \ref{fig:structure} shows the sequence of nuclides of the outer crust of a neutron star in $\beta$-equilibrium, taken from \cite{AME12} and in the case of $^{
82}$Zn from \cite{Wolf13b}. With the new mass value, the nuclide $^{82}$Zn is no longer present in the crust due to it being considerably less bound than predicted by HFB-19. The experimentally determined equilibrium composition of a cold, nonaccreting and non-rotating neutron star of 1.4 solar mass and 10\,km radius is now constrained to a new depth of 223\,m.

\section{Results and Discussion}
Extending our previous studies from 2006 \cite{Ruster06}, we are now in the position to investigate how different nuclear mass tables react in predicting the crustal composition in $\beta$-equilibrium using also new measurements of nuclear binding energies \cite{AME12} and in particular the new mass value of $^{82}$Zn.

To determine the EOS of the outer crust of neutron stars, it is necessary to identify for each given pressure the nucleus which minimizes the thermodynamic potential which is in this case the Gibbs free energy per baryon or baryon chemical potential. At $T=0$, the pressure and baryon chemical potential can be derived by standard thermodynamic relations from the total energy density $\epsilon_{\rm tot}$. In the most simple case, formulated by BPS, the energy density has only contributions from the mass of a given nucleus $W_N$, the Coulomb lattice energy $W_L$ and from electrons $\epsilon_e$:
\begin{equation}
\epsilon_{\rm tot}=n_N(W_N+W_L)+\epsilon_e \nonumber \; , 
\end{equation}
where $n_N$ is the number density of the nucleus. For a more accurate description, higher order corrections of the Coulomb and electron contribution can be implemented, like electron binding, screening or Thomas-Fermi, exchange, correlation or zero-point motion energy \cite{Salpeter61,Haensel94}. In the present calculations we include only the most important of these corrections, namely for electron binding, screening, and exchange energy in the same formulation as used in \cite{Pearson11} which was partly taken from \cite{Haensel94} who use results from \cite{Salpeter61} in the ultra-relativistic limit. For a selection of the mass models discussed below we provide data files of the detailed crustal composition and the EOS on one of the authors' personal homepage.\footnote{See
\texttt{http://phys-merger.physik.unibas.ch/{\raisebox{-0.25\baselineskip}{\textasciitilde}}hempel/eos.html}.}

\begin{figure}
\includegraphics[width=0.7\textwidth]{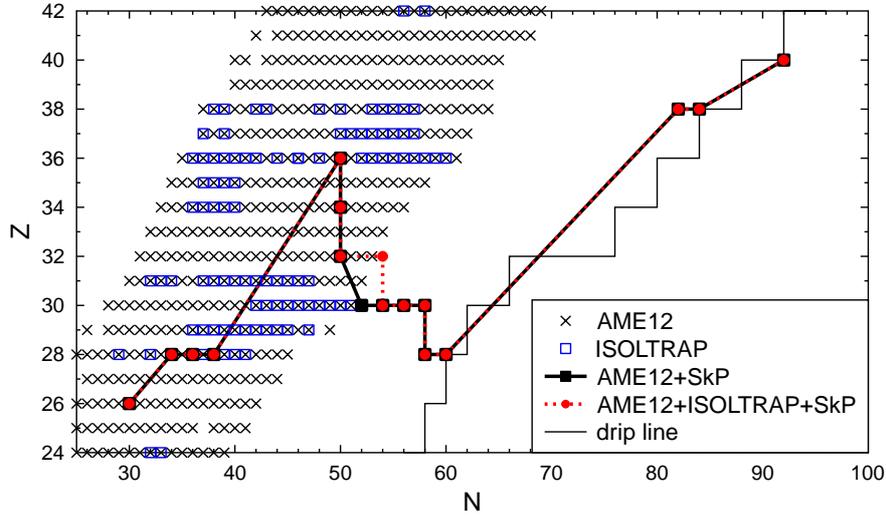}
\caption{\label{fig:sequence}Sequence of nuclides in the neutron-star outer crust for the SkP mass tables. In all calculations, the new mass value of $^{82}$Zn excludes the nucleus to be present in the crust. For details, see text.}
\end{figure}

With these crust calculations, the results from \cite{Wolf13b} for the three mass tables HFB-19, HFB-20, and HFB-21 could be reproduced. The only exception being $^{126}$Mo  which is predicted by HFB-19 in our calculations, but which occurs only in the narrow density interval $(2.09322 - 2.09399)\cdot10^{11}\,$g/cm$^3$. Furthermore, the location of the drip line differs as well, due to the different definition used here (and previously in [56]). We use an algorithm which determines the most asymmetric even-even nuclei with positive two-neutron separation energies and connect them by straight lines. Using this definition, all predicted nuclides lie in front of or on the dripline.

\begin{figure}
\includegraphics[width=0.6\textwidth]{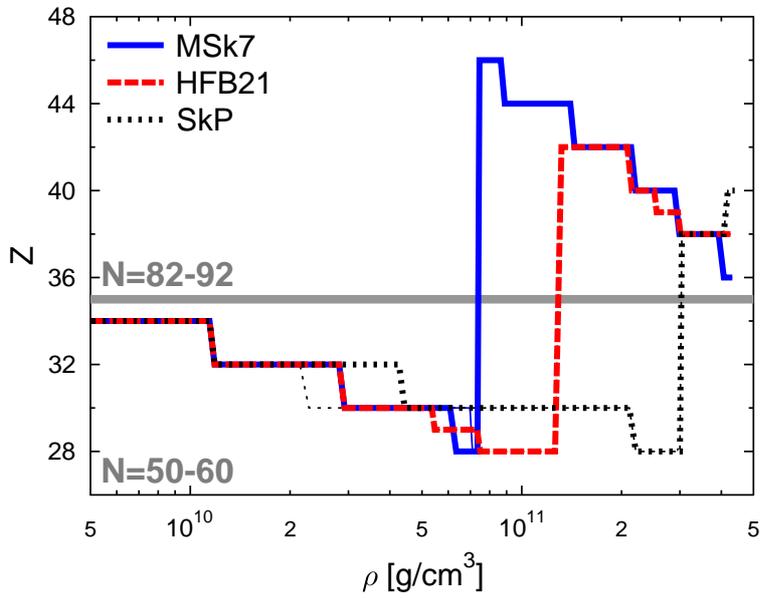}
\caption{\label{fig:zrho}Charge number as a function of density for three different mass models. Thick (thin) lines show the case with (without) considering the new mass value of $^{82}$Zn. For details, see text.}
\end{figure}

In a second step, other nuclear mass models which predicted $^{82}$Zn to be present in the crust, were re-evaluated. The sequence of nuclides in the outer crust has been calculated for the nuclear mass models HFB-8 \cite{Goriely03,Samyn04}, MSk7 \cite{Tondeur00,Goriely01}, and SkP \cite{Dobaczewski84,Stoitsov03,Dobaczewski04} (for details, see \cite{Ruster06,Guo07}). Since the mass of $^{82}$Zn is less bound than predicted, it is no longer present in the outer crust. As an example, Fig.~\ref{fig:sequence} shows the change in the sequence of nuclides for the mass model SkP mapped along the chart of nuclides. Further nuclides measured by ISOLTRAP and part of the sequence are marked by blue squares, the black crosses denote the AME2012 data base \cite{AME12}. The new sequence including the ISOLTRAP mass value of $^{82}$Zn is depicted by the red dashed line. It can also be seen from the plot that the last nucleus of the outer crust, i.e., before neutrons start to drip out, is located on the respective drip line 
of the model, which is true for all our calculations. In Fig.~\ref{fig:zrho} the sequence of nuclides of the outer crust of a neutron star for the mass model SkP is compared to the mass model MSk7 -- which also exhibits a change due to the new mass value --  and the mass model HFB-21, which has been discussed in \cite{Wolf13b} and which did not exhibit a change of the crustal composition. The charge number is plotted as a function of density, where the two regions of nuclides with magic neutron number $N \approx 50$ and $N \approx 82$ are delimited. None of the nuclides in the upper half of the graph have been measured yet. The thick (thin) lines show the sequence of nuclides with (without) the new mass value of $^{82}$Zn. Note, while the change in the sequence of nuclides is quite substantial for the mass model SkP, the mass model MSk7 exhibits only a slight change around a density of $7 \cdot 10^{10}$g/cm$^3$.

\begin{figure}
\includegraphics[width=0.9\textwidth]{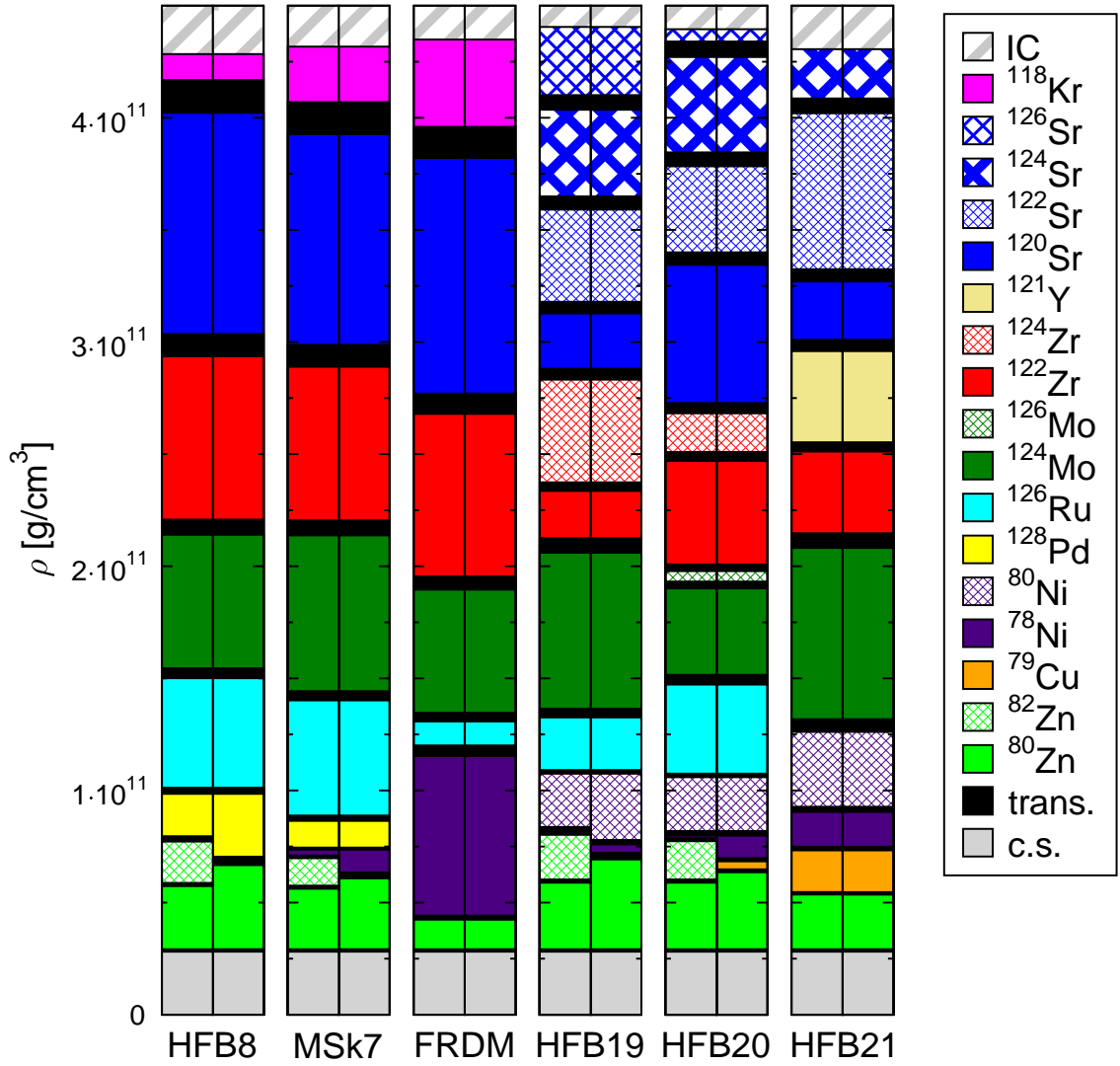}
\caption{\label{fig:bar_rho}The sequence of nuclides of different mass models with (right columns) and without (left columns) the new $^{82}$Zn mass. A certain color corresponds to a single element, where the density interval of the different isotopes is illustrated. The common sequence to all models is denoted by ``c.s.'', the last nucleus being $^{82}$Ge. The transition to the inner crust is marked by ``IC'', transitions between different nuclei by ``trans.''. For details, see text.}
\end{figure}

A set of 25 nuclear mass tables (the 21 mass tables used in \cite{Ruster06}, the mass model SLy6 of \cite{Guo07} and HFB19,20,21 of \cite{Goriely10}) were checked for the appearance of $^{82}$Zn, all of them showing that it is not a nucleus present in neutron stars. The new mass measurement discussed here further enhances the effect observed before \cite{Baym71b,Haensel89,Haensel94,Ruster06,Pearson12}: nuclides in the outer crust are predicted to have predominantly neutron numbers at the $N$ = 50 and 82 magic shells. In a previous paper \cite{Guo07} we confirmed the crucial role of nuclear shell effects in the chemical composition of neutron stars in $\beta$-equilibrium also for a mass model which includes triaxial nuclear deformations. So this result is quite robust. 

Finally, Fig.~\ref{fig:bar_rho} compares the findings of this work using the results from Wolf and colleagues \cite{Wolf13b} and the finite-range-droplet model \cite{Moller95}, a model which did not predict $^{82}$Zn from the start to be part of the crustal composition. We show the sequence of nuclides as a function of density including the new mass value of $^{82}$Zn. A certain color corresponds to a single element, where the density interval of the different isotopes is illustrated. If the color is combined with a pattern, the corresponding nucleus has a neutron number different from $N$=50 or $N$=82. Full colors without a pattern show nuclei having these neutron magic numbers. The common sequence to all models is denoted by ``c.s.'', the last nucleus being $^{82}$Ge. The transition to the inner crust is marked by ``IC''. The regions filled with black and marked with ``trans.'' show the transitions between different nuclei. Each of these transitions corresponds in our approach to a small first-order 
Maxwell phase transition. In a more sophisticated treatment different mixed lattices could be used \cite{Jog82}. In Fig.~\ref{fig:bar_rho} all our main results can be identified nicely: With the information of the new measurements, $^{82}$Zn is not expected to be present in the equilibrated crust of neutron stars any more. For most mass models, its place in the sequence of nuclei in the outer crust is replaced by nuclei with a $N=50$ neutron magic shell. The magic neutron shells $N=50$ and $N=82$ seem to be the dominating effect of nuclear structure regarding the crustal composition. We see that the transition to the inner crust seems to occur at about the same pressure for all models while the more recent microscopic models tend to predict a richer variety of species in the crust, with thinner layers.

Concerning higher order corrections for Coulomb and electron energies: we have found that in the bottom layers of the outer crust there can be changes in the sequence, namely that additional nuclides appear or that some are replaced, see also Ref.~\cite{Guo07}. This depends on which of these corrections are included and in which form. Even though the corrections are small, such a behavior is possible if different nuclides have similar Gibbs free energies in the minimization procedure. However, our results for the non-appearance of $^{82}$Zn were not influenced by these higher order corrections.

\section{Summary}
We have presented different ways in which nuclear masses influence our knowledge of neutron stars. The information available about these objects is based on astrophysical observations, mainly neutron-star mass data. Complementary to observational constraints are laboratory measurements, which place tight constraints on the range of nuclear parameters in the EOS. In light of recent mass measurements of $^{54}$Ca and $^{82}$Zn we have illustrated this link regarding advances in the description of neutron matter as well as the equilibrium composition of a neutron star's outer crust. Phenomenological or microscopic mass models usually produce a large scatter in their predictions for nuclear masses far from stability \cite{Lunney03}. The inclusion of three-body forces based on chiral EFT, which originate from a description of nuclear interactions using the symmetries of QCD, has led to the successful description of the calcium isotopic chain. Providing a successful testing ground for calculations using three-body 
forces in neutron-rich nuclides also strengthens the description of neutron-star matter. Chamel et al.~\cite{Chamel11} note that although several realistic calculations of the EOS of neutron-star matter all agree very closely at nuclear and sub-nuclear densities, they differ greatly in the predicted density dependence of the symmetry energy at the much higher densities towards the center of neutron stars. There are also very few data, either observational or experimental, to discriminate between the different possibilities. Chamel et al.~further conclude that fitting the nuclear force to the mass data is a necessary condition for obtaining reliable estimates of the masses of the unmeasured highly neutron-rich nuclides, see also \cite{Pearson13}. In this context, it is fascinating that ``an atom sheds light on neutron stars'' \cite{Grant13}. 

With the information of the new measurements, $^{82}$Zn is not expected to be present in the equilibrated crust of neutron stars any more. For most mass models, its place in the sequence of nuclei in the outer crust is replaced by nuclei with a $N=50$ neutron magic shell. The magic neutron shells $N=50$ and $N=82$ seem to be the dominating effect of nuclear structure regarding the crustal composition. The results presented in this work illustrate the fact that a single mass measurement influences the elemental composition of the neutron-star outer crust in $\beta$-equilibrium. This is in contrast to core-collapse r-process scenarios where it is difficult to identify how sensitive the r-process path is to a single measurement \cite{Xu13}. Some nuclear physics properties of nuclear matter can be studied in terrestrial laboratories with new neutron-rich radioactive beams. A possible mass measurement program is thus two-fold: to explore the limit of nuclear stability for enhancing our knowledge of the nuclear 
force as well as to determine the mass of certain, neutron-rich nuclides around the neutron shell closure $N=50$ and $N=82$ to probe the equilibrium composition of the outer crust of a neutron star. Microscopic approaches now allow self-consistent modeling of global neutron-star properties (i.e.~the EOS) together with the crustal composition (i.e.~binding energies). The predictions do not diverge from those obtained using binding energies of phenomenological mass models having many free parameters. Therefore, the latest theoretical advances, combined with those of mass spectrometry, contribute to robust and more confident neutron-star models. 

\section{Acknowledgments}
We thank K.~Hebeler for comments on the manuscript and R.~N.~Wolf for preparing Fig.~\ref{fig:structure}. SK thanks V.~Manea for discussions, and DL thanks S. Goriely, N. Chamel, and J.M. Pearson for communications. This work was supported by the German Federal Ministry for Education and Research (BMBF) (Grants No.~06GF9102 and No.~06GF9101I), the Max-Planck Society, the European Union seventh framework through ENSAR (Contract No.~262010), and the French IN2P3. MH is grateful for support from the Swiss National Science Foundation (SNF) under project number no.~200020-132816/1 and for participation in the EuroGENESIS collaborative research program of the European Science Foundation (ESF), in the ENSAR/THEXO project and in CompStar, a research networking program of the ESF.





\bibliographystyle{model1a-num-names}
\bibliography{Bibliography}







\end{document}